\documentclass[12pt]{article}
\usepackage{amssymb}
\usepackage{setspace}

\evensidemargin \oddsidemargin
\author{Latevi Lawson\\
Institut de Math\'ematiques et de Sciences Physiques (IMSP),\\
Laboratoire de Recherche en Physique Th\'eorique (LRPT),\\
01 BP 613 Porto-Novo, Rep. du B\'enin \\
Email: lawmenx@gmail.com
\\ \\
Laure Gouba \\
The Abdus Salam International Centre for
Theoretical Physics (ICTP),\\
 Strada Costiera 11,
I-34151 Trieste, Italy \\
Email: lgouba@ictp.it
\\ \\
Gabriel Y Avossevou \\
Institut de Math\'ematiques et de Sciences Physiques (IMSP),\\
Laboratoire de Recherche en Physique Th\'eorique (LRPT),\\ 
01 BP 613 Porto-Novo, Rep. du B\'enin \\
Email: gabriel.avossevou@imsp-uac.org}

\begin{title}
{Two-Dimensional Noncommutative Gravitational Quantum  Well}
\end{title}

\begin{document}

\maketitle

\abstract{ In this paper we consider two kinds of noncommutative space-time 
commutation relations in two-dimensional configuration space and feature the absolute 
value of the minimal length from the generalized uncertainty relations associated 
to the particular commutation relations.
We study the problem of the two-dimensional gravitational quantum well 
in the new Hermitian variables and confront
the experimental results for the first lowest 
energy state of the neutrons in the Earth's gravitational field to estimate 
the upper bounds on the noncommutativity parameters. The absolute value 
of the minimum length is smaller than a few nanometers.}

\section{Introduction}\label{sec1}

A new set of noncommutative space-time commutation 
relations in two-dimensional configuration space has been recently introduced \cite{fgs}.
The space-space commutation relations are deformations of the standard flat noncommutative 
space-time relations that have position dependent structure constants. 
These deformations lead to minimal lengths and it has been found that any object in this 
two dimensional space is string like in the 
sense that having a fundamental length in one direction beyond which a resolution is impossible. 
Under suitable transformations and choices new variables that are Hermitian have been set and 
some simple models have been solved in these new variables.  Some extensions of this work have 
been done in \cite{fgb, dfg} and the model of harmonic oscillator have been solved in these 
new variables \cite{alavi}.
The aim of the present work is to solve the analog of the problem of a particle in 
the quantum well of the Earth's gravitational field in the new variables obtained from deformed 
commutation relations.

The gravitational quantum well is obtained by considering a particle of mass $m$ moving in $xy$ plane 
subject to the Earth's gravitational field $\vec{g} = -g \vec{e}_x$ and a perfectly reflecting mirror  
placed at the bottom (at $x=0$). In ordinary quantum mechanics this problem is well known and studied
in text books \cite{goldman,landau,flugge,haar,sakurai}. A possibility for measuring 
such states for neutrons has been discussed in \cite{frank}. The lowest quantum state of neutrons 
in such a system has been observed experimentally at the Institut Laue-Langevin \cite{nesvi1,nesvi2}.
The existence of this phenomenon has been confirmed and studied in more details in \cite{nesvi3}.
These experiments gave the opportunities to confront observations and various theoretical models 
concerning quantum effects in gravity. Indeed various papers 
on gravitational quantum well in a noncommutative geometry can be found in the literature 
\cite{berto,rosa,brau,rabin,saha,lay,castello,dey}. 

In the present paper we consider the gravitational quantum well in two-dimensional quantum mechanics 
with deformed commutation relations and we determine how the parameters of deformation affect its energy 
spectrum and consider the experimental results to place upper bounds on the parameters of deformations. 
The paper is organized as follows. In the next section, we summarise the deformed commutation relations in 
two-dimensional quantum mechanics, the details being given in \cite{fgs}. In section (\ref{sec3}), we 
discuss the analog of the gravitational quantum well in the new variables obtained from the 
deformed commutation relations and determine for each case the shift in the energy spectrum. In the 
section (\ref{sec4}), we give concluding remarks where we compare our results with the existing results 
in the literature and study the upper bounds on the parameters of noncommutativity.

\section{Deformed commutation relations in two-dimensional quantum mechanics }\label{sec2}

A particle moving in d-dimensions is described in wave mechanics by a 
configuration space $\mathbb{R}^d$ and a Hilbert space $L^2$ of square integrable 
wavefunctions $\psi(x)$ over $\mathbb{R}^d$.  The inner product on $L^2$ is 
\begin{equation}
(\phi, \psi) = \int d^d x\phi^*(x)\psi(x).
\end{equation}
The elements of this Hilbert space are labelled by $\psi(x) \equiv \vert \psi \rangle$ 
and the elements of its dual by $\langle \psi \vert$, which maps elements of $L^2$ 
onto complex numbers by $\langle \phi\vert \psi\rangle =  (\phi, \psi)$. One of the basic 
axioms of commutative quantum mechanics is that physical observables correspond to Hermitian 
operators $A$ on $L^2$. 
In the two dimensional quantum system the  Heisenberg algebra is 
\begin{eqnarray}\label{eq1}\nonumber
{[\hat x_s,\hat y_s] = 0};\quad
{[\hat x_s, \hat p_{x_s} ]} =  i\hbar ;\quad
{[\hat y_s, \hat p_{y_s} ]} =  i\hbar;\\
{\quad [\hat x_s, \hat p_{y_s}]} = 0;\quad 
{\quad [\hat y_s, \hat p_{x_s}]} = 0;\quad
{[\hat p_{x_s},\hat p_{y_s}]} =  0,
\end{eqnarray}
where the operators $\hat x_s, \hat y_s, \hat p_{x_s},\hat p_{y_s}$ are 
Hermitian operators acting on the space of square integrable function.
A unitary representation of the algebra in (\ref{eq1})  is 
the Schr\"odinger representation 
\begin{eqnarray}
 \hat x_s\psi(x,y) &=& x \psi(x,y);\quad 
 \hat y_s\psi(x,y) = y \psi(x,y);\\
 \hat p_{x_s}\psi(x,y) &=& -i\hbar \frac{\partial}{\partial x}\psi(x,y);\quad 
 \hat p_{y_s}\psi(x,y) = -i\hbar \frac{\partial}{\partial y}\psi(x,y).
\end{eqnarray}

Two dimensional non-commutative 
quantum mechanics (NCQM) is based on a simple modification of the commutations relations 
between the hermitian position o\-pe\-ra\-tors $\hat x_0,\; \hat y_0$ and the hermitian momentum 
operators $\hat p_{x_0},\; \hat p_{y_0}$  which satisfy
\begin{equation}\label{eq2}
\begin{array}{ccc}
{[\hat x_0, \hat y_0] } =  i\theta; &
{[\hat x_0, \hat p_{x_0}]} = i\hbar; &
{[\hat y_0,\hat p_{y_0}]} = i\hbar; \\
{[\hat p_{x_0}, \hat p_{y_0} ]} = 0; &
{[\hat x_0, \hat p_{y_0}]} = 0; &
{[\hat y_0, \hat p_{x_0}]} = 0, 
\end{array}
\end{equation}
where $\theta \in \mathbb{R}$ is the spatial non-commutative parameter of 
dimension of length square. If $\theta$ is set to zero, we obtain the standard 
Heisenberg commutations relations (\ref{eq1}). 
The operators $\hat x_0,\hat y_0, \hat p_{x_0}, \hat p_{y_0}$  can be realized as follows 
\begin{eqnarray}\nonumber
\hat x_0\psi(x,y) = x\star \psi(x,y);  \quad 
\hat y_0\psi(x,y) = y\star \psi(x,y);\\
\hat p_{x_0}\psi(x,y) = -i\hbar\frac{\partial}{\partial x}\psi(x,y);\quad 
\hat p_{y_0}\psi(x,y) = -i\hbar\frac{\partial}{\partial y}\psi(x,y).
\end{eqnarray}
where $\star$ replaces the usual product of fields. The $\star$ product is defined as follows
\begin{equation}
 (f\star g )(x) = \exp \left(\frac{i}{2}\theta_{ij}\partial_{x_i}\partial_{y_j}\right) 
 f(x)g(y)\vert_{x=y}
\end{equation}
where $f$ and $g$ are two arbitrary infinitely differentiable functions on $\mathbb{R}^{3+1}$ and 
$\theta_{ij}$ is real and antisymmetric, ie., $\theta_{ij} = \theta\epsilon_{ij}$ 
( $\epsilon_{ij}$ a completely antisymmetric tensor with $\epsilon_{12} = 1$) .

We can express the phase space operators $\hat x_0, \hat y_0, \hat p_{x_0}, \hat p_{y_0}$ 
in terms of the standard phase space operators  $\hat x_s, \hat y_s, \hat p_{x_s}, \hat p_{y_s}$  
by performing the asymmetric Bopp-shift in space 
that can set in two different ways
\begin{equation}\label{bop1}
 \hat x_0 = \hat x_s - \frac{\theta}{\hbar} \hat p_{y_s}; \quad \hat y_0 = \hat y_s;
\end{equation}
or 
\begin{equation}\label{bop2}
 \hat x_0 = \hat x_s;\quad  \hat y_0 = \hat y_s + \frac{\theta}{\hbar} \hat p_{x_s};
\end{equation}
There are some advantages in using the asymmetric Bopp shift such as the decoupling of the 
variables in some of the problems and some simplifications of expressions. 
The equations (\ref{bop1}) and (\ref{bop2}) do not always lead to the same results for the same problems. 
For that reason the symmetrical Bopp shift 
\begin{equation}\label{bop3}
 \hat x_0 = \hat x_s -\frac{\theta}{(2\hbar)}\hat p_{y_s}; \quad 
 \hat y_0 = \hat y_s + \frac{\theta}{(2\hbar)}\hat p_{x_s}
\end{equation}
is often used. 

We consider new phase space operators $\hat x,\;\hat y,\;\hat p_x,\;\hat p_y$ that satisfy 
the following commutations 
relations
\begin{equation}\label{eq3}
 \begin{array}{ccc}
  {[\hat x,\hat y]} = i\theta (1+\tau \hat y^2); &  {[\hat x,\hat p_x]} = i\hbar(1 +\tau \hat y^2) & 
  {[\hat y,\hat p_y]} = i\hbar(1 +\tau \hat y^2); \\
  {[\hat p_x,\hat p_y] = 0}; & {[\hat x, \hat p_y]} = 2i\tau 
  \hat y (\theta \hat p_y + \hbar \hat x); & {[\hat y, \hat p_x] = 0}, 
 \end{array}
\end{equation}
where 
\begin{equation}\label{rep1}
\hat x =  \hat x_0 + i\theta\tau \hat y_0 + \tau \hat y_0^2\hat x_0; \quad \hat y = \hat y_0;\quad 
\hat p_x = \hat p_{x_0};\quad 
\hat p_y = \hat p_{y_0}  -i\hbar\tau \hat y_0  + \tau \hat y_0^2\hat p_{y_0},
\end{equation}
or by using the symmetric Bopp-Shift in equation (\ref{bop3})
\begin{eqnarray}\label{rep2}\nonumber
 \hat x & = & \hat x_s -\frac{\theta}{2\hbar}\hat p_{y_s} +\tau \hat y_s^2\hat x_s + \theta\tau\left(i\hat y_s +
 \frac{1}{\hbar}\hat y_s\hat p_{x_s}\hat x_s - \frac{1}{2\hbar}\hat y_s^2\hat p_{y_s}\right) \\\nonumber
  &+& \theta^2\tau\left( \frac{i}{2\hbar}\hat p_{y_s} +\frac{1}{(2\hbar)^2}\hat p_{x_s}^2 \hat x_s -
  \frac{1}{2\hbar^2}\hat y_s\hat p_{x_s}\hat p_{y_s}\right) -\frac{\tau\theta^3}{8\hbar^3}\hat p_{x_s}^2\hat p_{y_s};\\
 \hat y &=& \hat y_s + \frac{\theta}{2\hbar}\hat p_{x_s};\\\nonumber
 \hat p_x &=& \hat p_{x_s};\\\nonumber
 \hat p_y &=& \hat p_{y_s} + \tau (-i\hbar\hat y_s + \hat y_s^2 \hat p_{y_s}) + 
 \theta\tau\left(-\frac{i}{2}\hat p_{x_s} +\frac{1}{\hbar}\hat y_s\hat p_{x_s}\hat p_{y_s}\right) 
 + \frac{\tau\theta^2}{4\hbar^2}\hat p_{x_s}^2\hat p_{y_s}.
\end{eqnarray}
It is easy to verify that the variables $\hat x,\hat y,\hat p_x, \hat p_y $ are hermitian. 
The algebra (\ref{eq3}) is a position dependent deformed 
Heisenberg algebra and satisfy the Jacobi identity. 
We do not provide an analog of the Schr\"odinger representation for the variables 
$\hat x,\; \hat y, \; \hat p_x,\; \hat p_y $ but by mean of the  equations 
(\ref{rep1}) or (\ref{rep2}) one can define the action of those operators. 

The commutation relations in equations  (\ref{eq1}), (\ref{eq2}), (\ref{eq3} )
lead respectively to uncertainty relations. 

In the case of the equation (\ref{eq1}) we 
have the standard  uncertainty relations 
\begin{equation}
 \Delta \hat x_s\Delta \hat p_{x_s}\ge \frac{\hbar}{2}; \quad
 \Delta \hat y_s\Delta \hat p_{y_s}\ge \frac{\hbar}{2}.
\end{equation}

In the situation of the equation (\ref{eq2}), we have an additional uncertainty 
due to the noncommutativity of the position operators
\begin{equation}
 \Delta \hat x_0\Delta \hat y_0 \ge \frac{|\theta|}{2}; \quad 
 \Delta \hat x_0\Delta \hat p_{x_0}\ge \frac{\hbar}{2}; \quad
 \Delta \hat y_0\Delta \hat p_{y_0}\ge \frac{\hbar}{2}.
 \end{equation}
 
In the situation of the equation (\ref{eq3}), we feature the uncertainty relations 
in the positions operators 
\begin{equation}
 \Delta \hat x \Delta \hat y \ge \frac{1}{2}\vert \langle [\hat x, \hat y]\rangle\vert
\end{equation}
that is equivalent to 
\begin{eqnarray}
 \Delta \hat x \Delta \hat y \ge\frac{\theta}{2}(1 + \tau \langle \hat y ^2\rangle ),
\end{eqnarray}
we consider the fact that $\langle \hat y^2\rangle = \langle \hat y\rangle^2 + \Delta\hat y^2$ and 
we set the function 
\begin{equation}
 f(\Delta \hat x, \Delta \hat y) = \Delta\hat x\Delta\hat y - 
 \frac{\theta}{2}(1 +\tau\langle \hat y\rangle^2 + \tau\Delta\hat y^2)
\end{equation}
and solving 
\begin{equation}
 \partial_{\Delta\hat y}f(\Delta \hat x, \Delta \hat y)= 0 \quad \textrm{and} 
 \quad f(\Delta \hat x, \Delta \hat y) = 0, 
\end{equation}
leads to a minimal length in $\hat x$ in a simultaneous $\hat x,\;\hat y$ -measurement 
\begin{equation}\label{aml}
 \Delta \hat x_{\textrm{min}} = \theta\sqrt\tau\sqrt{1 +\tau\langle \hat y \rangle ^2}\;,
\end{equation}
the absolute minimal value $\Delta \hat x_{\textrm{min}}$ being $\theta\sqrt{\tau}$.

\section{Deformed commutation relations and the gra\-vi\-ta\-tional quantum well }\label{sec3}

We consider in two-dimensional configuration space, the model of a particle of mass $m$ 
subjected to the Earth's gravitational field in one direction that is the vertical taken to 
be described by $(-g)\vec{e}_x$. In the direction transverse to the gravitational field, $y$, 
the particle is free.  The Hamiltonian of the system is given by 
\begin{equation}\label{wham1}
 \hat H_0 = \frac{1}{2m}\hat p_{x_s}^2 + \frac{1}{2m}\hat p_{y_s}^2 + m g \hat x_s,
\end{equation}
where the operators $\hat x_s,\; \hat y_s,\; \hat p_{x_s},\;\hat p_{y_s}$ satisfy the 
commutation relations in equation (\ref{eq1}).
The Schr\"odinger equation is given by 
\begin{equation}\label{sh1}
 \hat H_0 \Psi(x,y) = E \Psi(x,y).
\end{equation}
As it is clearly seen, the system is decoupled and the solution to the 
eigenvalue equation (\ref{sh1}) is given by 
\begin{equation}
 \Psi(x,y) = \psi_n(x)\psi(y),\quad E\equiv E_{n,k};
\end{equation}
where $\psi(y)$ is the wave function in the y-direction and $\psi_n(x)$ the 
wave function in the x-direction.
Since the particle is free in the y-direction, the wave function is 
\begin{equation}
 \psi(y)= \int_{-\infty}^{+\infty}dk g(k)e^{iky},
\end{equation}
where $g(k)$ determines the shape of the wave packet and the energy spectrum is continuous
\begin{equation}
 E_k = \frac{\hbar^2k^2}{2m}\,.
\end{equation}
In the x-direction, the eigenfunctions can be expressed in terms of Airy function $\phi(z)$
 with appropriate normalization \cite{landau} as follows
\begin{equation}
 \psi_n(x) = \alpha_n\phi(z),
\end{equation}
where the normalization factor for the n-th eigenstate is given by: 
\begin{equation}
 \alpha_n = \left( \left(\frac{\hbar^2}{2m^2g}\right)^{\frac{1}{3}}\int_{r_n}^{+\infty}dz\phi^2(z)\right)^{-\frac{1}{2}};\quad 
 z = \left(\frac{2m^2g}{\hbar^2}\right)^{\frac{1}{3}} \left(x -\frac{E_n}{mg}  \right)\,,
\end{equation}
and the eigenvalues are determined by the roots $r_n$ of the Airy function as follows
\begin{equation}
 E_n = -\left( \frac{mg^2\hbar^2}{2}\right)^{\frac{1}{3}} r_n\,. 
\end{equation}
The spectrum of the system is then 
\begin{equation}
 \Psi(x,y)= \psi_n(x)\psi(y), \quad E_{n,k} = -\left(\frac{mg^2\hbar^2}{2}\right)^{\frac{1}{3}}\times r_n + \frac{\hbar^2 k^2}{2 m}.
\end{equation}

In this paragraph we consider the analog of the quantum gravitational well on noncommutative flat space. 
The Hamiltonian of the system is given by
 \begin{equation}\label{wham2}
 \hat H = \frac{1}{2m}\hat p_{x_0}^2 + \frac{1}{2m}\hat p_{y_0}^2 + m g \hat x_0 ,
\end{equation}
where the operators $\hat x_0,\;\hat y_0,\; \hat p_{x_0},\;\hat p_{y_0}$ satisfy the 
algebra (\ref{eq2}). By mean of the symmetric Bopp shift in equation (\ref{bop3}), we can rewrite
the Hamiltonian in equation (\ref{wham2}) as follows
\begin{equation}
\hat H_\theta = \frac{1}{2m}\hat p_{x_s}^2 + \frac{1}{2m}\hat p_{y_s}^2 + m g (\hat x_s -\frac{\theta}{2\hbar}\hat p_{y_s}),
\end{equation}
then 
\begin{equation}\label{tham2}
 \hat H_\theta = \frac{1}{2m}\hat p_{x_s}^2 + \frac{1}{2m}\hat p_{y_s}^2 + m g \hat x_s -\theta\frac{mg}{2\hbar} \hat p_{y_s},
\end{equation}
that is equivalent to 
\begin{equation}\label{nt}
 \hat H_\theta = \hat H_0  -\theta\frac{mg}{2\hbar} \hat p_{y_s}.
\end{equation}
We consider $\theta$ positive and very small and therefore must be a small correction at 
the low-energy level. We treat then the new term appearing in (\ref{nt}) as 
a perturbation in the commutative Hamiltonian $H_0$. The shift caused by this term 
on the system's energy levels is given by the expection value of the perturbation on 
the system's wave function that is 
 \begin{eqnarray}\nonumber
 \Delta E_{n,k} &=& - \theta\frac{ m g}{2\hbar}\langle \Psi(x,y) \vert \hat p_{y_s}\vert \Psi(x,y)\rangle;\\
  &=& - \theta\frac{ m g k}{2} \;, 
\end{eqnarray}
and 
\begin{equation}
 E_{n,k,\theta} = -\left(\frac{m g^2\hbar^2}{2}\right)^{\frac{1}{3}}\times r_n 
 + \frac{\hbar^2 k^2}{2m} -\theta\frac{ m g k}{2} .
\end{equation}

Let us consider now the analog of the quantum gravitational 
well from position-dependent noncommutativity. 
The Hamiltonian is given by
\begin{equation}\label{foa}
 \hat H = \frac{1}{2m}\hat p_x^2 + \frac{1}{2m}\hat p_y^2  + m g\hat x, 
\end{equation}
where the operators $\hat x,\,\hat p_x,\, \hat p_y$ satify the commutation relations in equation (\ref{eq3}).  
In order to find the spectrum one can work directly in the position dependent noncommutative variables 
or by mean of the represention in equation (\ref{rep1}) or the representation in equation (\ref{rep2}). 
The problem is that in a situation of noncommutativity of the spatial operators there is an ambiguity 
in the meaning of the wavefunctions in the position representation.
So we study the problem using the representation in equation (\ref{rep2}) where the operators 
$\hat x,\hat y, \hat p_x,\hat p_y$ are expressed in terms of the operators $\hat x_s,\hat y_s, \hat p_{x_s}, 
\hat p_{y_s}$. We assume that the parameters $\tau$ and $\theta$ are positive and very small and only 
the leading approximation in $\theta$ and $\tau$ is considered. The terms proportional to 
$\hat y^2_s$ and $\hat p_{y_s}^2$ do not affect the particle's energy spectrum in the direction 
of the gravitational field \cite{berto}, the expectation value $\langle \hat y_s\rangle$ is a finite value 
which is expected for a localized group of plane waves and will not produce any shift on the energy levels \cite{berto}. 
We set the expectation value $<\hat y_s >$ to be zero in order to ensure the hermiticity condition 
of the operator $\hat p_y$ in equation (\ref{rep2}) . The Hamiltonian (\ref{foa}) is then reduced to
\begin{equation}
 \hat H_{\theta,\tau} =  \hat H_0 - \theta\frac{m g}{2\hbar}\hat p_{y_s }
  - \tau\frac{\hbar^2}{2m}.
\end{equation}
 The shift being determined by 
\begin{eqnarray}
 \Delta E_{n,k} &=& \langle \Psi(x,y)\vert \left( - \theta\frac{m g}{2\hbar}\hat p_{y_s } - 
 \tau\frac{\hbar^2}{2m}\right)
 \vert \Psi(x,y)\rangle \,;\\
&=&- \theta\frac{m g k}{2} -\tau\frac{\hbar^2}{2m}\,,
\end{eqnarray}
so
\begin{equation}
 E_{n,k,\theta,\tau} = -\left(\frac{m g^2\hbar^2}{2}\right)^{\frac{1}{3}}\times r_n + 
 \frac{\hbar^2 k^2}{2m} -\theta\frac{ m g k}{2} -\tau\frac{\hbar^2}{2m}.
\end{equation}

\section{Concluding remarks}\label{sec4}

We consider in this work two kinds of deformed commutation relations in two-dimensional 
quantum mechanics. The first kind being the most used 
in noncommutative quantum mechanics since it leads to less ambiguities. 
The second kind is dynamical and leads to 
the existence of minimal length. 

For the first kind, we study the analog of the gravitational quantum well in the new variables 
induced by the deformation parameter $\theta$. The energy of the system is given by
\begin{equation}
 E_{n,k,\theta} = -\left(\frac{m g^2\hbar^2}{2}\right)^{1/3} r_n + \frac{\hbar^2 k^2}{2m} -\frac{\theta m g}{2}k . 
\end{equation}
The first remark is that the energy shift caused by this deformed commutation relations is negative. 
The noncommutativity parameter $\theta$ affects the energy spectrum and that 
has been also stated in \cite{castello}. In a situation of noncommutativity in both 
configuration and momentum spaces it has been found that to the 
leading order noncommutativity in configuration space does not affect the energy spectrum of 
the system \cite{berto}. Confronting our theoretical results to 
the experiments done in references \cite{nesvi1,nesvi2}, 
in the sense that the absolute value of the shift induced by the noncommutativity parameter $\theta$ 
should be smaller than the maximum diffence of the energy levels provided by the experiments, 
we have for the first level energy ($n=1$)
\begin{equation}\label{in1}
 \theta\frac{m g k}{2} \le \Delta E_1^{exp},
\end{equation}
with 
\begin{equation}
 \Delta E_1^{exp} = 6.55\times 10 ^{-32}\; J .
\end{equation}
For the neutrons 
 $\textrm{m} = 1.675\times 10^{-27}$ kg, and in the experiments the neutrons 
 had a mean horizontal speed $\langle v_y\rangle = 6.5m/s$ so that 
 $k = \langle p_y \rangle /\hbar = m\langle v_y\rangle/\hbar = 1.03\times 10^8\;\textrm{m}^{-1}$. 
 Then by considering 
 $\textrm{g} = 9.81\; \textrm{m}\textrm{s}^{-2}$ and 
 $\hbar = 10.59\times 10^{-35} \textrm{Js}$, we can calculate 
the upper bound for the parameter $\theta$ from equation (\ref{in1}) as
\begin{equation}\label{b1}
 \theta \lesssim 0.755 \times 10^{-13}\; \textrm{m}^2 .
\end{equation}
The upper bounds on the parameter of spatial noncommutativity $\theta$ have been also established in 
 \cite{castello, carol,jabbari, irina}: in \cite{castello} where
 the experimental context is the same with the present paper, the upper bound establised is 
 $\theta \lesssim 0.771 \times 10^{-13}\;\textrm{m}^2$ for $n=1$, in \cite{carol} 
 existing experiments related to Lorentz invariance bound the parameter of the spatial
 noncommutativity to $\theta \lesssim 0.388 \times 10^{-33}\;\textrm{m}^2 $,
 in \cite{jabbari} measurements of the Lamb shift establish the upper bound to $\theta \lesssim 0.388\times 10^{-39}\;\textrm{m}^2$,
 in \cite{irina} the upper bound  $\theta \lesssim 0.155\times 10^{-54}\; \textrm{m}^2$ is established from 
an analysis of clock-comparison experiments. The upper bounds of the parameter of spatial noncommutativity $\theta$ depend 
on the experiments and in order to improve the results more accurate experimental data are needed.

In the second case, we study the dynamical deformed commutation relations that 
is a deformation of the commutation relations of flat noncommutative space-time. 
This kind of deformed commutation relations lead to the existence of minimal length, the 
absolute minimal value being $\Delta\hat x_{\textrm{min}}= \theta\sqrt{\tau}$ m.
The spectrum of the Hamiltonian is given by
 \begin{equation}
 E_{n,k,\theta,\tau} = - \left(\frac{m g^2\hbar^2}{2}\right)^{1/3} r_n + \frac{\hbar^2}{2m}k^2 
 -\frac{\theta m g}{2}k -\tau\frac{\hbar^2}{2m}\,.
\end{equation}
Here again the energy shift caused by this deformed commutation relations is negative.
In order to confront with the experimental results we compare the absolute value of the energy shift with the
experiments results at the first level, that is 
\begin{equation}\label{in2}
 \theta\frac{ m g k}{2} + \tau\frac{\hbar^2}{2m}\le \Delta E_1^{exp},\;\; \textrm{with}\;\;
 \Delta E_1^{exp} = 6.55\times 10 ^{-32}\; J.
\end{equation}
Here, appears the effects of both parameters of noncommutativity $\tau$ and $\theta$. 
 In order to be consistent with the experiments the parameters $\theta$ and $\tau$ should 
 satisfy the inequality
\begin{equation}\label{in4}
 8.46 \times 10^{-19}\;\theta + 3.34 \times 10^{-42}\;\tau \lesssim 6.55\times 10^{-32}.
\end{equation}
From equation (\ref{b1}) we can estimate the upper bound for the parameter $\tau$ to be
\begin{equation}\label{b2}
 \tau \lesssim 6.26\times 10^8\, \textrm{m}^{-2}.
\end{equation}
In our assumption $\tau$ is considered to be very small and  therefore 
the bound of $\tau$ in equation (\ref{b2}) is not a good estimation for 
the parameter $\tau$.

With respect to equation (\ref{b1}) and equation (\ref{b2}), 
the absolute value of the minimal length is bounded as 
\begin{equation}\label{ba3}
 \Delta \hat x_{\textrm{min}} = \theta\sqrt{\tau}\, \lesssim \,1.87 \times 10^{-9}\, \textrm{m}.
\end{equation}
The upper bound of the absolute value of the minimal length in equation (\ref{ba3}) 
confirms the conclusion in \cite{brau}. Indeed, F. Brau and F. Buisseret studied the 
dynamics of a particle in a gravitational quantum well in the context of 
nonrelativistic quantum mechanics with a two dimensional analog of the modified Heisenberg 
algebra $[\hat x,\hat p] = i (1+\beta^2 \hat p^2),\; \hbar = c = 1$. They conclude that 
if $\beta$ is a quantity that depends on the energetic content of the system an upper bound 
can be derived from the experimental results \cite{nesvi1,nesvi2} and the minimum length scale 
associated to neutrons moving in  a gravitational quantum well is smaller than a 
few nanometers. \\
 
\noindent {\bf Acknowledgments}: L. L. is supported by Institut de Math\'ematiques et de Sciences Physiques (IMSP)
in Porto-Novo, Rep. Benin under grants from the Abdus Salam International Centre for Theoretical Physics (ICTP)-
Trieste/Italy and the German Academic Exchange Service (DAAD). L. G would like to acknowledge support from 
the Abdus Salam International Centre for Theoretical Physics (ICTP)-Trieste/Italy.


\begin{thebibliography}{30}

\bibitem{fgs}{Andreas Fring, Laure Gouba and Frederik G. Scholtz, Strings from position-dependent noncommutativity,
 J. Phys. A: Math. Theor. 43 (2010) 345401}
\bibitem{fgb}{Andreas Fring, Laure Gouba, Bijan Bagchi, Minimal areas from q-deformed oscillator algebras, 
J. Phys. A: Math. Theor. 43 (2010) 425202 }
\bibitem{dfg}{Sanjib Dey, Andreas Fring, Laure Gouba, $\mathcal{PT}$-symmetric noncommutative spaces 
with minimal volume uncertainty relations,
J. Phys. A: Math. Theor. 45 (2012) 385302}
\bibitem{alavi}{ S. A. Alavi and S. Abbaspour, Dynamical noncommutative quantum mechanics, 
 J. Phys. A: Math. Theor. 47 (2014) 045303}
\bibitem{goldman}{I. I. Goldman, V. D. Krivchenkov, V. I. Kogan, V. M. Galitscii, Problems in Quantum Mechanics (New York, Academic, 1960)}
\bibitem{landau}{L. D. Landau, E. M Lifshitz, Quantum Mechanics (Oxford, Pergamon, 1976)}
\bibitem{flugge}{S. Fl\"ugge, Practical Quantum Mechanics I (Berlin, Springer, 1974)}
\bibitem{haar}{D. ter Haar, Selected Problems in Quantum Mechanics (Academic, New York, 1964)}
\bibitem{sakurai}{J. J. Sakurai, Modern Quantum Mechanics (Benjamin/Cummings, Menlo Park, 1985)}
\bibitem{frank}{V. I. Luschikov, A. I. Frank, Quantum effects occuring when ultracold neutrons 
are stored on a plane, Pis' ma Zh. Eksp. Teor. Fiz. {\bf 28} No 9, 607-609 (5 November 1978)}
\bibitem{nesvi1}{ V. V. Nesvizhevsky, H. G. B\"orner, A. K. Petukhov, H.
Abele, S. Bae\ss ler, F. J. Rue\ss, Th. St\"oferle, A. Westphal,
A. M. Gagarsky, G. A. Petrov, A. V. Strelkov, Quantum states of neutrons in the Earth's gravitational field,
Nature 415, 297-299 (2002) }
\bibitem{nesvi2}{V. V. Nesvizhevsky, H. G. B\"orner, A. M. Gagarski, A. K.
Petukhov, G. A. Petrov, H. Abele, S. Bae\ss ler, G. Divkovic,
F. J. Rue\ss, Th. St\"oferle, A. Westphal, A. V. Strelkov, K. V.
Protasov, A. Yu. Voronin, Measurement of quantum states of neutrons in the Earth's gravitational field, 
Phys. Rev. D {\bf 67}, 102002 (2003)}
\bibitem{nesvi3}{V. V. Nesvizhevsky, A. K. Petukhov, H. G. B\"orner, T. A. Baranova, A. M. Gagarski, G. A. Petrov, K. V. Protasov,
A. Yu. Voronin, S. Bae\ss ler, H. Abele, A. Westphal, L. Lucovac, Study of the neutron quantum states in the gravity field, 
Eur. Phys. J. C 40: 479 - 491, 2005}
\bibitem{berto}{O. Bertolami, J. G. Rosa, C. M. L. de Arag$\tilde{a}$o, P. Castorina, 
D. Zappal\`a, Noncommutative Gravitational Quantum Well, 
Phys. Rev. D {\bf 72} 025010 (2005)}
\bibitem{rosa}{O. Bertolami, J. G. Rosa, The Gravitational Quantum Well, J. Phys. Conf. Ser. 33 (2006) 118-130}
\bibitem{brau}{Fabian Brau, Fabian Buisseret, Minimal length uncertainty relation and gravitational quantum well, 
Phys. Rev. D {\bf 74}, 0360002 (2006)}
\bibitem{rabin}{Rabin Banerjee, Binayak Dutta Roy and Saurav Samanta, Remarks on the noncommutative gravitational quantum well, 
 Phys. Rev. D {\bf74} 045015 (2006)}
\bibitem{saha}{Anirban Saha, Time-Space Noncommutativity in Gravitational Quantum Well scenario, Eur. Phys. J. C51:199-205, 2007 }
\bibitem{lay}{ Lay Nam Chang, Zachary Lewis, Djordje Minic, and Tatsu Takeuchi, 
On the Minimal Length Uncertainty Relation and the Foundations of String Theory, Advances in High Energy Physics 2011, 493514 (2011)}
\bibitem{castello}{K. H. C. Castello-Branco and A. G. Martins, Free-fall in a unifrom gravitational field in non-commutative quantum 
mechanics, J. Math. Phys. 51:102106, 2010 }
\bibitem{dey}{Anha Bhat, Sanjib Dey, Mir Faizal, Chenguang Hou, Qin Zhao, Modification of Schr\"odinger-Newton equation 
due to braneworld models with minimal length,
 Phys. Lett. B 770 (2017) 325-330 }
\bibitem{carol}{Sean M. Carroll, Jeffrey A. Harvey, V. Alan Kosteleck\'y, Charles D. Lane, and Takemi. Okamoto, 
Noncommutative Field Theory and Lorentz Violation, Phys. Rev. Lett. {\bf 87} (2001) 141601}
\bibitem{jabbari}{M. Chaichian, M. M. Sheikh-Jabbari, A. Tureanu, Hydrogen Atom Spectrum and the Lamb Shift in Noncommutative QED, 
Phys. Rev. Lett. {\bf 86} (2001) 2716 }
\bibitem{irina}{Irina Mocioiu, Maxim Pospelov, Radu Roiban, Low-energy limits on the antisymmetric tensor 
field background on the brane and on the non-commutative scale, Phys. Lett. B 489 (2000) 390 -396 }
\end{thebibliography}
\end{document}